\documentclass[twocolumn]{aastex63}

\usepackage{amssymb,amsmath,amsthm}
\usepackage{graphicx}
\usepackage{mathrsfs}
\usepackage{booktabs}
\usepackage{multirow}
\usepackage{empheq}
\usepackage{verbatim}
\usepackage{mathtools}
\usepackage{hyperref}
\usepackage{xcolor}
\usepackage{xspace}

\newcommand{\dmc}{DMC\xspace}
\newcommand{\themis}{\textsc{Themis}\xspace}
\newcommand{\uas}{$\mu$as\xspace}
\newcommand{\snr}{signal-to-noise ratio\xspace}

\defcitealias{TMS}{TMS}
\defcitealias{Palumbo_2020}{PWP}

\begin{document}

\title{A D-term Modeling Code (DMC) for simultaneous calibration and full-Stokes imaging of very long baseline interferometric data}

\correspondingauthor{Dominic~W.~Pesce}
\email{dpesce@cfa.harvard.edu}

\author[0000-0002-5278-9221]{Dominic~W.~Pesce}
\affiliation{Center for Astrophysics $|$ Harvard \& Smithsonian, 60 Garden Street, Cambridge, MA 02138, USA}
\affiliation{Black Hole Initiative at Harvard University, 20 Garden Street, Cambridge, MA 02138, USA}

\begin{abstract}
In this paper we present \dmc, a model and associated tool for polarimetric imaging of very long baseline interferometry datasets that simultaneously reconstructs the full-Stokes emission structure along with the station-based gain and leakage calibration terms.  \dmc formulates the imaging problem in terms of posterior exploration, which is achieved using Hamiltonian Monte Carlo sampling.  The resulting posterior distribution provides a natural quantification of uncertainty in both the image structure and in the data calibration.  We run \dmc on both synthetic and real datasets, the results of which demonstrate its ability to accurately recover both the image structure and calibration quantities as well as to assess their corresponding uncertainties.  The framework underpinning \dmc is flexible, and its specific implementation is under continued development.
\end{abstract}

\keywords{keywords}

\section{Introduction}

Interferometric observations in radio astronomy natively access the so-called ``visibility domain,'' with each visibility determined by the complex correlation between the electric fields incident at a pair of telescopes \citep[][hereafter \citetalias{TMS}]{TMS}.  These visibilities provide information about the Fourier transform of the incident flux distribution via the van Cittert-Zernike theorem, and radio interferometric imaging -- i.e., the process by which the visibility measurements are translated into a sky-plane image -- presents an example of an ill-posed inverse problem.  The combination of sparse Fourier-plane sampling and uncertain calibration, both of which are exacerbated for very long baseline interferometric (VLBI) observations, prevents a direct inversion of the visibilities to produce a unique image.  Instead, images must be ``reconstructed'' with the aid of additional assumptions about the image structure (e.g., flux positivity, source sparsity) to overcome this non-uniqueness.

A variety of algorithms exist for reconstructing images in radio interferometry, and these algorithms can be broadly classified into the two categories established in \cite{Paper4}.  ``Inverse modeling'' schemes, exemplified by the CLEAN algorithm and its variants \citep{Hogbom_1974,Clark_1980,Schwab_1984}, operate directly with the inverse Fourier transform of the visibility measurements and seek to iteratively deconvolve the effects of the finite sampling from the reconstructed image.  ``Forward modeling'' schemes, such as the maximum entropy \citep[e.g.,][]{Nityananda_1982,Cornwell_1985} and regularized maximum likelihood \citep[e.g.,][]{Chael_2016,Akiyama_2017} methods, instead parameterize the image structure (typically using a grid of pixels) and Fourier transform it to predict the values of the visibility measurements.  The image parameters are then varied so as to optimize some objective function, typically consisting of a data comparison term (e.g., a $\chi^2$ metric) along with one or more regularization terms.

From the perspective of computational speed, the CLEAN approach has historically been a clear favorite for VLBI imaging.  The forward modeling schemes, though generally more computationally taxing, benefit from the ability to enforce various nonlinear constraints (such as flux positivity) on the image and to fit directly to non-visibility data products \citep[such as closure quantities; e.g.,][]{Chael_2018}.  Typical implementations of both classes of algorithm, however, share a mixed relationship with data calibration whereby image reconstruction steps are iterated with interleaving ``self-calibration'' steps that attempt to solve for station-based calibration terms \citep[e.g.,][]{Readhead_1980}.  Furthermore, both the inverse and forward modeling classes of image reconstruction algorithm classically lack a natural quantification of uncertainty in the image.

Recent developments have yielded a new class of image reconstruction algorithms, based on posterior exploration or parameterization techniques, that aim to overcome the aforementioned shortcomings \citep[e.g.,][]{Cai_2018a,Cai_2018b,Arras_2019,Broderick_2020b}.  From the perspective of statistical integrity, an image reconstruction algorithm should solve simultaneously for the ensemble of both sky-plane emission structures and requisite calibration terms that are permissible, given the uncertainties in the data and any sources of prior knowledge about the parameters.  In this paper we present such an algorithm in terms of a model for simultaneous calibration and full-Stokes imaging of VLBI data, along with an implementation of this model within a generic posterior exploration framework.  An implementation of an analogous model within \themis \citep{Broderick_2020a} is presented in a separate paper, \citet{Broderick_2021}.

This paper is organized as follows.  In \autoref{sec:ModelSpecs} we provide the a detailed description of the model, specify our likelihood construction, and describe its software implementation.  In \autoref{sec:Demonstrations} we demonstrate the results of fitting this model to both synthetic and real VLBI datasets.  We summarize and conclude in \autoref{sec:Summary}.

\section{Model specifications} \label{sec:ModelSpecs}

For the compact sources and small fields of view typically considered in VLBI, the observed visibilities are related to the Fourier transform of the sky brightness distribution by the van Cittert-Zernike theorem \citepalias{TMS},

\begin{equation}
\tilde{I}(u,v) = \iint I(x,y) e^{-2 \pi i (u x + v y)} dx dy ,
\end{equation}

\noindent where we use a tilde (~$\tilde{}$~) to denote a transformed quantity.  In this section and throughout the paper, unless otherwise specified, we consider observations made with only a single frequency channel.

We have developed a new publicly available D-term Modeling Code (\dmc)\footnote{\url{https://github.com/dpesce/eht-dmc}} that implements the polarimetric image model detailed in this section.  \dmc is implemented in Python, and it fits the model using a Bayesian formalism in which the posterior distribution $\mathcal{P}(\boldsymbol{\Theta})$ of the parameter vector $\boldsymbol{\Theta}$ is related to the likelihood $\mathcal{L}(\boldsymbol{\Theta})$ and prior $\pi(\boldsymbol{\Theta})$ via Bayes' Theorem,

\begin{equation}
\mathcal{P}(\boldsymbol{\Theta}) \propto \mathcal{L}(\boldsymbol{\Theta}) \pi(\boldsymbol{\Theta}) .
\end{equation}

\noindent Model parameters and their associated priors are aggregated in \autoref{tab:ModelParameters}.  \dmc uses the \texttt{eht-imaging} library \citep{Chael_2016,Chael_2018} for internal organization and manipulation of VLBI data, and it uses the PyMC3 library \citep{Salvatier_2016} for sampling.  Because it uses a Markov chain Monte Carlo (MCMC) sampler, the output of running \dmc on a VLBI dataset is an ensemble of images that are drawn from the posterior distribution of the model.  From this ensemble it is possible to compute various useful statistics (e.g., means, variances), which we demonstrate in \autoref{sec:Demonstrations}.

\begin{deluxetable*}{LcC}[t]
\tablecolumns{3}
\tablewidth{0pt}
\tablecaption{Model parameters and priors \label{tab:ModelParameters}}
\tablehead{\colhead{Parameter} & \colhead{Description} & \colhead{Default prior}}
\startdata
F & image-integrated Stokes I flux density & \mathcal{N}_0(\breve{F},[0.1 \breve{F}]^2) \\
I_j / F & fraction of Stokes I flux density contained in pixel $j$ & \text{Dir}(N_{\text{pix}},\boldsymbol{1}) \\
p_j & polarization fraction in pixel $j$ & \mathcal{U}(0,1) \\
\alpha_j & azimuthal angle Poincar\'e coordinate in pixel $j$ & \mathcal{U}_{\text{per}}\left( -\pi, \pi \right) \\
\beta_j & polar angle Poincar\'e coordinate in pixel $j$ & \cos(\beta_j) \sim \mathcal{U}\left( -1, 1 \right) \\
\Sigma & FWHM of Gaussian blurring kernel, in $\mu$as & \delta(\Sigma-\breve{\Sigma}) \\
x_0 & overall image centroid shift along the Right Ascension (RA) axis, in $\mu$as  & \delta(x_0-\breve{x}_0) \\
y_0 & overall image centroid shift along the Declination (Dec) axis, in $\mu$as & \delta(y_0-\breve{y}_0) \\
\midrule
g_{R,a} & righthand gain amplitude for station $a$ & \mathcal{N}_0(1,\breve{\sigma}_{R,a}^2) \\
\theta_{R,a} & righthand gain phase for station $a$ & \mathcal{U}_{\text{per}}\left( -\pi, \pi \right) \\
g_{L,a} & lefthand gain amplitude for station $a$ & \mathcal{N}_0(1,\breve{\sigma}_{L,a}^2) \\
\theta_{L,a} & lefthand gain phase for station $a$ & \mathcal{U}_{\text{per}}\left( -\pi, \pi \right) \\
d_{R,a} & righthand leakage amplitude for station $a$ & \mathcal{U}(0,1) \\
\delta_{R,a} & righthand leakage phase for station $a$ & \mathcal{U}_{\text{per}}\left( -\pi, \pi \right) \\
d_{L,a} & lefthand leakage amplitude for station $a$ & \mathcal{U}(0,1) \\
\delta_{L,a} & lefthand leakage phase for station $a$ & \mathcal{U}_{\text{per}}\left( -\pi, \pi \right) \\
f & fractional systematic uncertainty & \mathcal{U}(0,1) \\
\midrule
N_x & number of image pixels along the RA axis & \ldots \\
N_y & number of image pixels along the Dec axis & \ldots \\
\text{FOV}_x & field of view along the RA axis, in $\mu$as & \ldots \\
\text{FOV}_y & field of view along the Dec axis, in $\mu$as & \ldots \\
\enddata
\tablecomments{A list of the model parameters and their corresponding prior distributions.  The top portion of the table lists the parameters associated with the image, the middle portion lists parameters associated with the calibration, and the bottom portion lists the hyperparameters.  We use a breve (~$\breve{}$~) to denote user-specified quantities.  We use a number of different prior classes: $\delta(x-a)$ denotes a Dirac delta prior over $x$ such that it takes on the fixed value $a$, $\mathcal{U}(a,b)$ denotes a uniform prior on the range $[a,b]$; $\mathcal{U}_{\text{per}}(a,b)$ denotes a periodic (or ``wrapped'') uniform prior on the range $[a,b]$; $\mathcal{N}(\mu,\sigma^2)$ denotes a normal (Gaussian) distribution with mean $\mu$ and variance $\sigma^2$; $\mathcal{N}_0(\mu,\sigma^2)$ denotes a normal distribution (with mean $\mu$ and variance $\sigma^2$) that has a lower-bound truncation at zero; $\mathcal{N}_c(\mu,\sigma^2)$ denotes a circularly-symmetric complex normal distribution with (complex) mean $\mu$ and variance $\sigma^2$ along both the real and imaginary directions; $\text{Dir}(N,\boldsymbol{a})$ denotes a Dirichlet distribution in $N$ dimensions with concentration parameter vector $\boldsymbol{a}$.}
\end{deluxetable*}

\subsection{Image model} \label{sec:Image}

We model the image as a Cartesian grid of $N_{\text{pix}}$ pixels, with the grid axes aligned with the equatorial coordinate axes.  Each pixel has a location $(x_j,y_j)$ and a Stokes I intensity $I_j$.  These intensities are constrained to sum to a total flux density $F$,

\begin{equation}
F = \sum_{j=1}^{N_{\text{pix}}} I_j , \label{eqn:ConstrainedSum}
\end{equation}

\noindent with $F$ specifiable but by default sampled from a normal prior truncated at zero to ensure positivity.  The constrained sum in \autoref{eqn:ConstrainedSum} is imposed via a Dirichlet prior on the pixel intensity values,

\begin{equation}
\frac{\boldsymbol{I}}{F} \sim \text{Dir}(N_{\text{pix}},\boldsymbol{a}) ,
\end{equation}

\noindent where $\boldsymbol{I} = \left( I_1, I_2, \ldots, I_{N_{\text{pix}}} \right)$ is the vector of pixel intensities.  The concentration parameter vector $\boldsymbol{a}$ is specifiable but defaults to $\boldsymbol{a} = \boldsymbol{1} \equiv \left( 1, 1, \ldots, 1 \right)$, which corresponds to a flat Dirichlet prior with the flux-normalized pixel intensities sampled uniformly on the ($N_{\text{pix}}-1$)-dimensional simplex.  Setting smaller values for the concentration parameters encourages sparsity in the image, while setting larger values encourages diffuse flux. \\

In each pixel, we can relate the Stokes I intensity $I_j$ to the other Stokes parameters by the inequality

\begin{equation}
I_j^2 \geq Q_j^2 + U_j^2 + V_j^2 \equiv p_j^2 I_j^2 ,
\end{equation}

\noindent where we have introduced the polarization fraction $p_j \leq 1$.  This spherical relationship lends itself naturally to a Poincar\'e parameterization in terms of angular variables,

\begin{equation}
\begin{pmatrix}
I_j \\
Q_j \\
U_j \\
V_j
\end{pmatrix} = I_j \begin{pmatrix}
1 \\
p_j \cos(\alpha_j) \sin(\beta_j) \\
p_j \sin(\alpha_j) \sin(\beta_j) \\
p_j \cos(\beta_j)
\end{pmatrix} , \label{eqn:Poincare}
\end{equation}

\noindent where $-\pi \leq \alpha_j \leq \pi$ is an azimuthal angle and $0 \leq \beta_j \leq \pi$ is a polar angle.  The angle $\alpha_j$ determines the orientation of the polarization ellipse, and it is related to the usual electric vector position angle (EVPA) $\chi_j$ by

\begin{equation}
\chi_j = \frac{1}{2} \tan^{-1}\left( \frac{U_j}{Q_j} \right) = \frac{\alpha_j}{2} .
\end{equation}

\noindent The angle $\beta_j$ determines the degree of circular polarization, with purely linear polarization having $\beta_j = \pi/2$ and purely circular polarization having $\beta_j = 0$ or $\beta_j = \pi$.  We sample $p_j$ from a unit uniform distribution, and we sample the angular variables uniformly on the unit sphere.  Our polarized image model thus consists of the four quantities $(I_j, p_j, \alpha_j, \beta_j)$ for every pixel, which together with the total flux $F$ amount to $4N_{\text{pix}}$ model parameters.

From the parameters $(I_j, p_j, \alpha_j, \beta_j)$, we determine the Stokes parameters in each pixel using \autoref{eqn:Poincare}.  We then compute the Fourier transforms of these Stokes parameters via

\begin{equation}
\begin{pmatrix}
\tilde{I}_k \\
\tilde{Q}_k \\
\tilde{U}_k \\
\tilde{V}_k
\end{pmatrix} = S_k \sum_{j=1}^{N_{\text{pix}}} A_{jk} \begin{pmatrix}
I_j \\
Q_j \\
U_j \\
V_j
\end{pmatrix} ,
\end{equation}

\noindent where

\begin{equation}
A_{jk} = \exp\left( -2 \pi i [u_k (x_j - x_0) + v_k (y_j - y_0)] \right)
\end{equation}

\noindent are elements of the discrete Fourier transform matrix, $(u_k,v_k)$ are the Fourier-plane coordinates for visibility measurement $k$ in units of the observing wavelength, $(x_0,y_0)$ are the image-plane coordinates of the image origin (or ``phase center''), and

\begin{equation}
S_k = \exp\left[ - \frac{\pi^2 \Sigma^2 (u_k^2 + v_k^2)}{4 \ln(2)} \right] \label{eqn:Smoothing}
\end{equation}

\noindent is a circularly symmetric Gaussian smoothing kernel with full width at half maximum (FWHM) $\Sigma$ in the image plane that serves to maintain image continuity.

\subsection{Corruption model} \label{sec:Corruptions}

For an array observing with circularly polarized feeds, the measured quantities are parallel- and cross-hand correlation products; we denote the parallel-hand visibilities as $RR_{12} \equiv \langle E_{R,1} E_{R,2}^* \rangle$ and $LL_{12} \equiv \langle E_{L,1} E_{L,2}^* \rangle$, and we denote the cross-hand visibilities as $RL_{12} \equiv \langle E_{R,1} E_{L,2}^* \rangle$ and $LR_{12} \equiv \langle E_{L,1} E_{R,2}^* \rangle$.  The measured visibilities on a baseline $ab$ are related to the Stokes visibilities on that same baseline by

\begin{equation}
\begin{pmatrix}
RR_{ab} \\
LL_{ab} \\
RL_{ab} \\
LR_{ab}
\end{pmatrix} = \begin{pmatrix}
\tilde{I}_{ab} + \tilde{V}_{ab} \\
\tilde{I}_{ab} - \tilde{V}_{ab} \\
\tilde{Q}_{ab} + i \tilde{U}_{ab} \\
\tilde{Q}_{ab} - i \tilde{U}_{ab}
\end{pmatrix} .
\end{equation}

In real interferometric observations, the measured visibilities are corrupted by a combination of a priori unknown signal propagation effects.  Following the radio interferometer measurement equation (RIME) formalism developed by \cite{Hamaker_1996} -- and in particular the $2 \times 2$ matrix extension described in \cite{Hamaker_2000} -- we relate the incident and measured visibilities using Jones matrix transformations of the ``coherency matrix,''\footnote{This particular (sky-intrinsic) coherency matrix is also referred to in the literature as the ``brightness matrix,'' and its measured counterpart has been referred to as the ``visibility matrix'' \citep{Smirnov_2011}.}

\begin{equation}
\mathbf{V}_{ab} \equiv \begin{pmatrix}
RR_{ab} & RL_{ab} \\
LR_{ab} & LL_{ab}
\end{pmatrix} .
\end{equation}

\noindent Within the RIME formalism, the (complex) Jones matrix $\mathbf{J}_a$ captures all linear transformations undergone by the incident astrophysical signal at a station $a$, such that

\begin{equation}
\hat{\mathbf{V}}_{ab} = \mathbf{J}_a \mathbf{V}_{ab} \mathbf{J}_b^{\dag} , \label{eqn:RIME}
\end{equation}

\noindent where a dagger (~$^{\dag}$~) denotes a conjugate transpose and a hat (~$\hat{}$~) denotes an observed quantity.

\dmc incorporates a minimal but standard \citepalias[see, e.g.,][]{TMS} threefold decomposition of $\mathbf{J}_a$,

\begin{equation}
\mathbf{J}_a = \mathbf{G}_a \mathbf{D}_a \mathbf{F}_a , \label{eqn:JonesMatrix}
\end{equation}

\noindent where

\begin{equation}
\mathbf{G}_a = \begin{pmatrix}
G_{R,a} & 0 \\
0 & G_{L,a}
\end{pmatrix}
\end{equation}

\noindent contains the station gain terms,

\begin{equation}
\mathbf{D}_a = \begin{pmatrix}
1 & D_{R,a} \\
D_{L,a} & 1
\end{pmatrix}
\end{equation}

\noindent contains the polarimetric leakage terms, and

\begin{equation}
\mathbf{F}_a = \begin{pmatrix}
e^{-i \phi_a} & 0 \\
0 & e^{i \phi_a} 
\end{pmatrix}
\end{equation}

\noindent applies the feed rotation angle, $\phi_a$.  Note that this decomposition incorporates only station-based corruptions, and it does not account for direction-dependent effects \citep[e.g.,][]{Smirnov_2011} or for other baseline-based corruptions.  The feed rotation angle $\phi_a$ depends on the station mount properties and on the source parallactic and elevation angles as a function of time, but for most radio interferometers it is well-known a priori; we thus assume the $\phi_a$ to be given and therefore do not incorporate them as model parameters.  The station gain and leakage terms, however, are typically imperfectly calibrated and so we retain both as model parameters.

We parameterize the complex station gains using amplitude and phase,

\begin{subequations}
\begin{equation}
G_{R,a} = g_{R,a} e^{i \theta_{R,a}} ,
\end{equation}
\begin{equation}
G_{L,a} = g_{L,a} e^{i \theta_{L,a}} .
\end{equation}
\end{subequations}

\noindent Our priors on the gain amplitudes are normal with a lower-bound truncation at zero, and we impose periodic uniform\footnote{A periodic uniform distribution is one that is uniformly distributed on the unit circle.} priors on the gain phases with the range $(-\pi,\pi)$.  For each observation we select a single ``reference station'' for which both the right and left gain phases are fixed to be zero (i.e., $\theta_{R} = \theta_{L}$).\footnote{Note that if the reference station does not actually have a zero-valued phase difference between its right- and left-hand gains, this treatment will result in an overall image EVPA rotation that must be absolutely calibrated \citep{Brown_1989}.}  We permit all gains other than those of the reference station to be independent across stations and across timestamps.  We note that in real-world arrays, the station gain amplitudes are not expected to fluctuate wildly from one timestamp to the next.  In this sense, \dmc aims to provide a conservative treatment of the gain beharior; i.e., \dmc permits -- though it does not impose -- large gain amplitude fluctuations between timestamps, such as may occur when a telescope is re-pointed.

We use an analogous parameterization for the complex leakage terms,

\begin{subequations}
\begin{equation}
D_{R,a} = d_{R,a} e^{i \delta_{R,a}} ,
\end{equation}
\begin{equation}
D_{L,a} = d_{L,a} e^{i \delta_{L,a}} .
\end{equation}
\end{subequations}

\noindent We impose unit uniform priors on the leakage amplitudes and periodic uniform priors on the leakage phases with the range $(-\pi,\pi)$. We assume that the leakage terms are constant in time \citep[see, e.g.,][]{Conway_1969,Roberts_1994} and so assign only a single $\mathbf{D}_a$ for every station.

We note that the default priors described in this section for the gain and leakage terms may be overridden when running \dmc to incorporate any a priori knowledge of the station properties.

\begin{figure*}
\centering
\includegraphics[width=\textwidth]{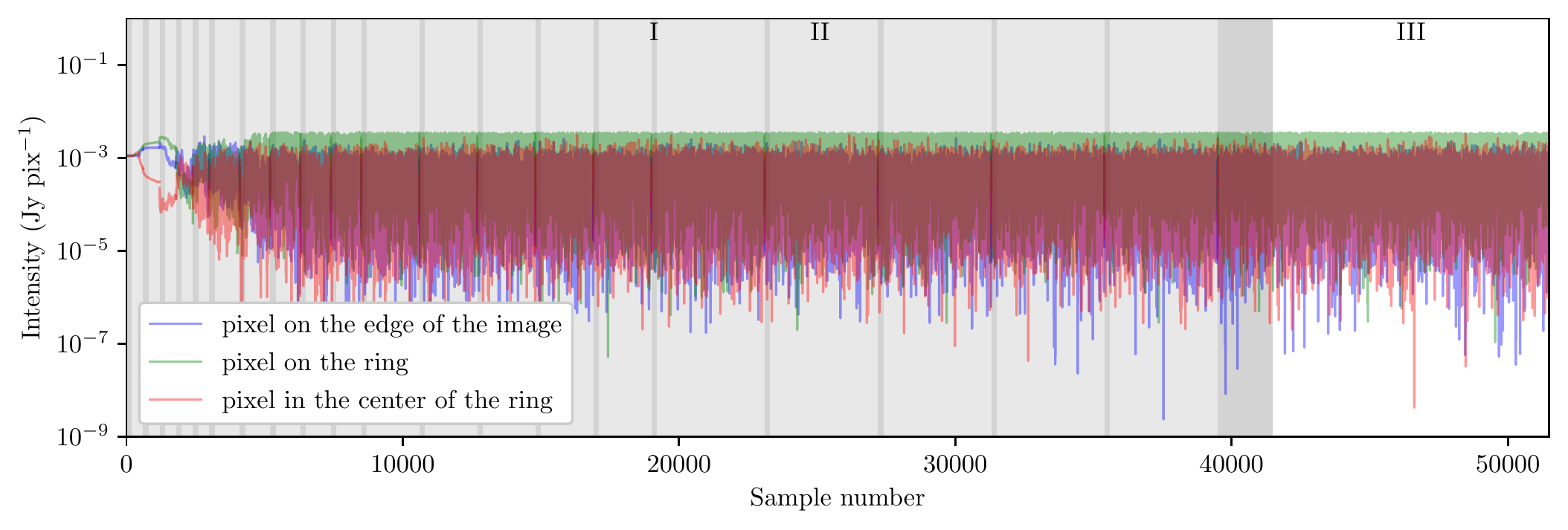}
\caption{Example set of parameter traces from a \dmc fit to the EHT-like synthetic dataset described in \autoref{sec:SyntheticData}, showing the Stokes I intensity of a pixel located towards the edge of the image (in blue), a pixel situated on the ring (in green), and a pixel located in the center of the ring (in red).  The shaded regions highlight the different tuning windows described in \autoref{sec:Tuning}; the dark gray shading indicates a ``fast'' tuning window (one example is labeled as ``I''), the light gray shading indicates a ``slow'' tuning window (one example is labeled as ``II''), and the unshaded region indicates the main sampling phase (labeled as ``III'').}
\label{fig:tuning}
\end{figure*}

\subsection{Likelihood construction}

The thermal noise in any single visibility measurement depends on various factors, including the collecting area of the telescopes constituting the baseline and the averaging time and bandwidth of the observation, but for the vast majority of sources of interest\footnote{The ``self-noise'' of \cite{Kulkarni_1989} introduces statistical dependence in the noise measured across multiple baselines, but this contribution only becomes relevant for extremely bright sources.} this thermal noise is normally distributed and statistically independent across different baselines \citepalias{TMS}.  In the absence of any other corruptions, a measured visibility is drawn from a circularly symmetric complex Normal distribution, e.g.,

\begin{equation}
\hat{RR}_{k} \sim \mathcal{N}_c\left( \mathcal{RR}_{k}, \sigma_{\text{th},RR,k}^2 \right) ,
\end{equation}

\noindent where $\mathcal{RR}_{k}$ is the ``true'' visibility value on baseline $k$ and $\sigma_{\text{th},RR,k}^2$ is the thermal variance in the corresponding visibility measurement.  Note that the presence of gain corruptions does introduce covariance between visibility measurements, but by explicitly modeling these gains we account for this covariance and ensure that the remaining differences between modeled and observed visibilities will be independently distributed \citep{Blackburn_2020}.

Though we explicitly incorporate a number of known corrupting effects into the model (see \autoref{sec:Corruptions}), we also permit an additional multiplicative systematic noise component,

\begin{equation}
\begin{pmatrix}
\sigma_{RR,k}^2 \\
\sigma_{LL,k}^2 \\
\sigma_{RL,k}^2 \\
\sigma_{LR,k}^2
\end{pmatrix} = \begin{pmatrix}
\sigma_{\text{th},RR,k}^2 + f^2 I_{k}^2 \\
\sigma_{\text{th},LL,k}^2 + f^2 I_{k}^2 \\
\sigma_{\text{th},RL,k}^2 + f^2 I_{k}^2 \\
\sigma_{\text{th},LR,k}^2 + f^2 I_{k}^2
\end{pmatrix} , \label{eqn:SystematicNoise}
\end{equation}

\noindent where $f$ is sampled from a unit uniform distribution.  This systematic component aims to account for any uncalibrated non-closing errors that cannot be described by thermal noise or leakage.  We then construct the likelihood of a particular set of model visibilities given the visibility measurements using

\scriptsize
\begin{subequations}
\begin{equation}
\mathcal{L}_{RR} = \prod_k \frac{1}{2 \pi \sigma_{RR,k}^2} \exp\left[ \frac{\left( \hat{RR}_k - RR_k \right) \left( RR_k - \hat{RR}_k \right)^*}{2 \sigma_{RR,k}^2} \right] ,
\end{equation}
\begin{equation}
\mathcal{L}_{LL} = \prod_k \frac{1}{2 \pi \sigma_{LL,k}^2} \exp\left[ \frac{\left( \hat{LL}_k - LL_k \right) \left( LL_k - \hat{LL}_k \right)^*}{2 \sigma_{LL,k}^2} \right] ,
\end{equation}
\begin{equation}
\mathcal{L}_{RL} = \prod_k \frac{1}{2 \pi \sigma_{RL,k}^2} \exp\left[ \frac{\left( \hat{RL}_k - RL_k \right) \left( RL_k - \hat{RL}_k \right)^*}{2 \sigma_{RL,k}^2} \right] ,
\end{equation}
\begin{equation}
\mathcal{L}_{LR} = \prod_k \frac{1}{2 \pi \sigma_{LR,k}^2} \exp\left[ \frac{\left( \hat{LR}_k - LR_k \right) \left( LR_k - \hat{LR}_k \right)^*}{2 \sigma_{LR,k}^2} \right] ,
\end{equation}
\end{subequations}
\normalsize

\noindent where the products are taken over all visibility measurements.  The final likelihood expression is then simply the product of the individual visibility likelihoods,

\begin{equation}
\mathcal{L} = \mathcal{L}_{RR} \mathcal{L}_{LL} \mathcal{L}_{RL} \mathcal{L}_{LR} .
\end{equation}

\subsection{Sampler and tuning} \label{sec:Tuning}

\dmc uses the Hamiltonian Monte Carlo (HMC; \citealt{Duane_1987}) No U-Turn Sampler (NUTS; \citealt{Hoffman_2011}) implemented within the PyMC3 Python package \citep{Salvatier_2016} to explore the posterior space.  HMC is an MCMC method whose output product is an ensemble of samples from the posterior distribution. Detailed descriptions of the HMC method can be found in, e.g., \cite{Neal_2011} and \cite{Betancourt_2017}.  PyMC3 is a probabilistic programming tool that leverages Theano \citep{Bergstra_2010,Bastien_2012} to automatically differentiate the posterior density when computing model gradients.

As an HMC sampler, PyMC3 exploits model gradient information to efficiently explore the high-dimensional posterior space presented by the polarized image model.  The number of tunable hyperparameters is minimized through the use of NUTS, but there remain two key hyperparameters that need to be adaptively tuned during sampling itself: a ``step size'' hyperparameter that sets the discretization interval for trajectory integrations, and a ``mass matrix'' hyperparameter (actually a collection of hyperparameters, the elements of the matrix) that determine the Gaussian distribution from which the momentum parameters are sampled.  PyMC3 natively adapts the step size hyperparameter during sampling \cite[see][]{Hoffman_2011}, but its default functionality only adapts the diagonal elements of the mass matrix.  Strong correlations in the posterior distribution can therefore lead to decreased sampling efficiency.

To mitigate this potential deficiency, we divide the sampling period into multiple tuning windows during which both the step size and (dense) mass matrix are adaptively determined.  We have designed these windows to mimic the ``warmup epochs'' used in the Stan package \citep{Carpenter_2017}.  An initial ``fast'' window is used to tune the step size parameter, after which a series of increasingly heavily-sampled ``slow'' windows are used to estimate the mass matrix using the parameter covariances measured from the set of posterior samples in the previous window.  Each slow window is preceded by a brief fast window to permit the step size to adapt to the new mass matrix.  A final fast window follows the last slow window, after which the main sampling phase proceeds using the tuned values for both hyperparameters.  This tuning procedure is illustrated in \autoref{fig:tuning}.

In practice, we find that tuning of both the step size and the mass matrix is essential for polarized imaging using \dmc.  If, e.g., the mass matrix tuning is restricted to only the diagonal elements, parameter autocorrelation times are liable to increase by several orders of magnitude and the sampler will effectively stall.

\section{Demonstrations} \label{sec:Demonstrations}

In this section we demonstrate the imaging capabilities of \dmc on both synthetic and real data.

\subsection{Synthetic data construction} \label{sec:SyntheticData}

\begin{figure}
\centering
\includegraphics[width=\columnwidth]{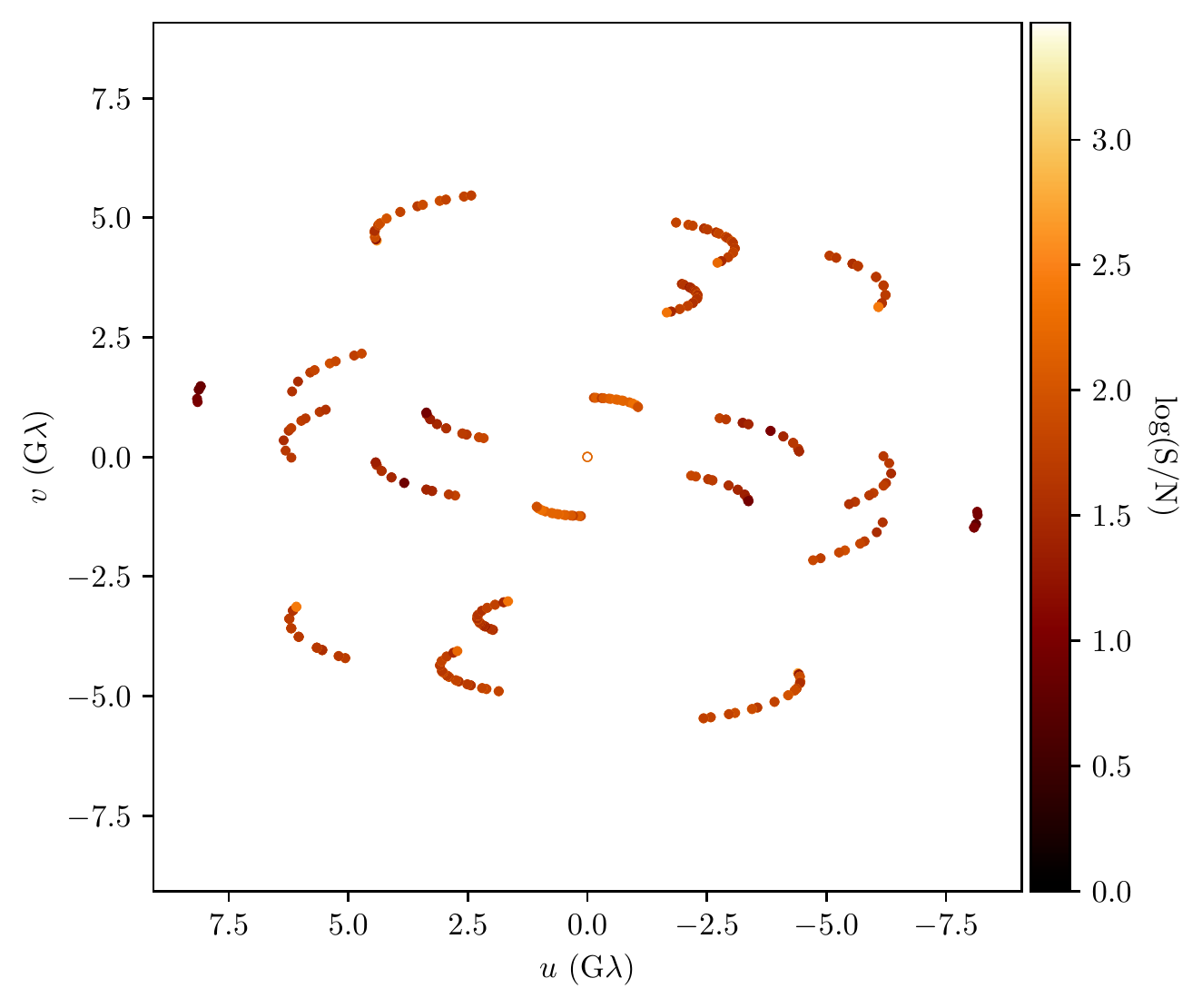}
\caption{$(u,v)$-coverage for the EHT-like synthetic dataset, with points colored by the base-10 logarithm of their Stokes I \snr.}
\label{fig:uv_coverage}
\end{figure}

\begin{figure*}
\centering
\includegraphics[width=\textwidth]{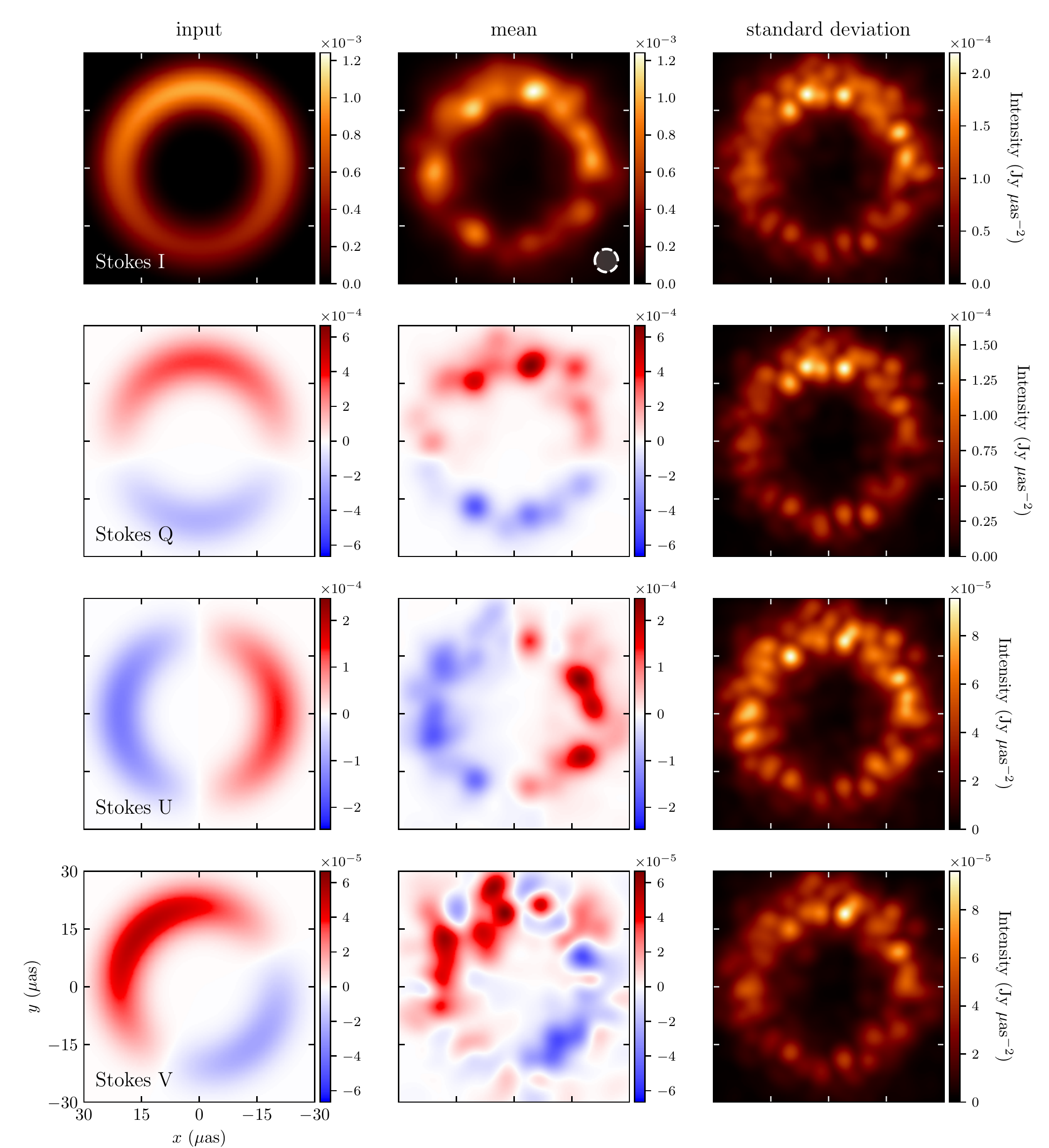}
\caption{\dmc image reconstructions of the polarized EHT-like synthetic dataset described in \autoref{sec:SyntheticData}.   Each row shows the results for a different Stokes parameter.  The same field-of-view is used for all plots and is explicitly labeled in the bottom-left panel.  The leftmost column shows the ground-truth input images, the middle column shows the mean posterior images for each of the Stokes parameters, and the rightmost column shows the standard deviations of the image posteriors.  Because these datasets included arbitrary gain phase corruptions, absolute position information is not uniquely recovered and so the reconstructed images have been shifted to the location that maximizes the normalized cross-correlation between the ground-truth Stokes I image and the mean of the Stokes I image posterior.  The 6\,\uas FWHM of the Gaussian smoothing kernel (i.e., $\Sigma$ from \autoref{eqn:Smoothing}) is shown in the bottom right-hand corner of the top middle panel.  Note that this kernel does not represent a typical ``restoring beam'' that gets applied after imaging has been completed; rather, it is a convolving function that gets self-consistently applied during the imaging process itself (see \autoref{sec:Image}).}
\label{fig:synthetic_DMC}
\end{figure*}

We first run \dmc on a synthetic dataset constructed to have properties similar to the 2017 observations of the M87 black hole with the EHT \citep{Paper1,Paper2,Paper3,Paper4,Paper5,Paper6}.  The baseline coverage and \snr distribution for this dataset is shown in \autoref{fig:uv_coverage}, and the input source model Stokes images are shown in the left panels of \autoref{fig:synthetic_DMC}.  The visibility data are generated in a circular polarization basis, corresponding to the state of the real EHT data after fringe-fitting has been performed \citep{Paper3}.

\begin{figure*}
\centering
\includegraphics[width=\textwidth]{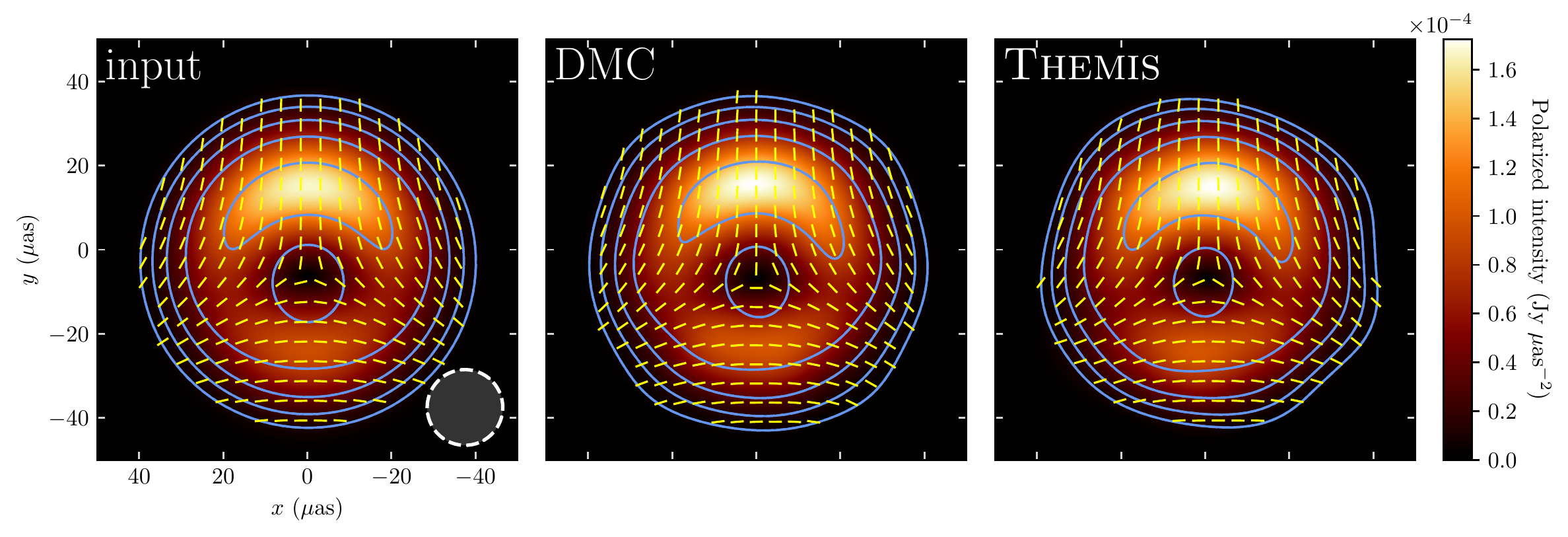}
\caption{A comparison of the linear polarization structure in the EHT-like dataset as recovered using \dmc (center) and \themis (right), with the input image shown in the leftmost panel.  The Stokes I structure is marked using blue contours, with the outermost contour levels starting at 5\% of the peak intensity and increasing inwards by factors of 2.  The background colormap indicates the linearly polarized intensity, and the overlaid tick marks show the EVPA direction.  All three panels share a common field of view (explicitly labeled in the leftmost panel), and all three images have been convolved with the EHT beam (circular Gaussian of width 18\,\uas; \citealt{Paper4}) shown in the lower right-hand corner of the left panel.}
\label{fig:EHT_pol_comparison}
\end{figure*}

The input source structure is a circular crescent with a diameter of 40\,\uas and a Gaussian FWHM of 10\,\uas, oriented such that the brightest region of the crescent is located towards the North.  The image-integrated flux density is 1.0\,Jy, and it is polarized at the 10\% level in linear polarization and at the 2\% level in circular polarization.  We construct the linear polarization structure to have an approximately threefold azimuthal symmetry (see the left panel of \autoref{fig:EHT_pol_comparison}), corresponding to nonzero $\beta_{-1}$ and $\beta_1$ modes in the decomposition developed by \citet[][hereafter \citetalias{Palumbo_2020}]{Palumbo_2020}.

In addition to the thermal noise, we add gain and leakage corruptions to each of the seven stations in the synthetic dataset.  The gain amplitudes are permitted to vary at the 10\% level, while the gain phases are unconstrained (i.e., they are sampled uniformly on the unit circle); each station has independent gains, but these gains are drawn from the same distribution and thus share the same magnitude of fluctuations. The same gain corruptions are used for both right- and left-hand polarization.  Each station has an associated complex station leakage in both right- and left-hand polarization that is at the level of 0--10\%.  The station gain amplitudes and phases are independently generated for each station at each observing timestamp, while the complex right- and left-hand leakage terms are held constant for each station across the synthetic observation.

\begin{figure}
\centering
\includegraphics[width=\columnwidth]{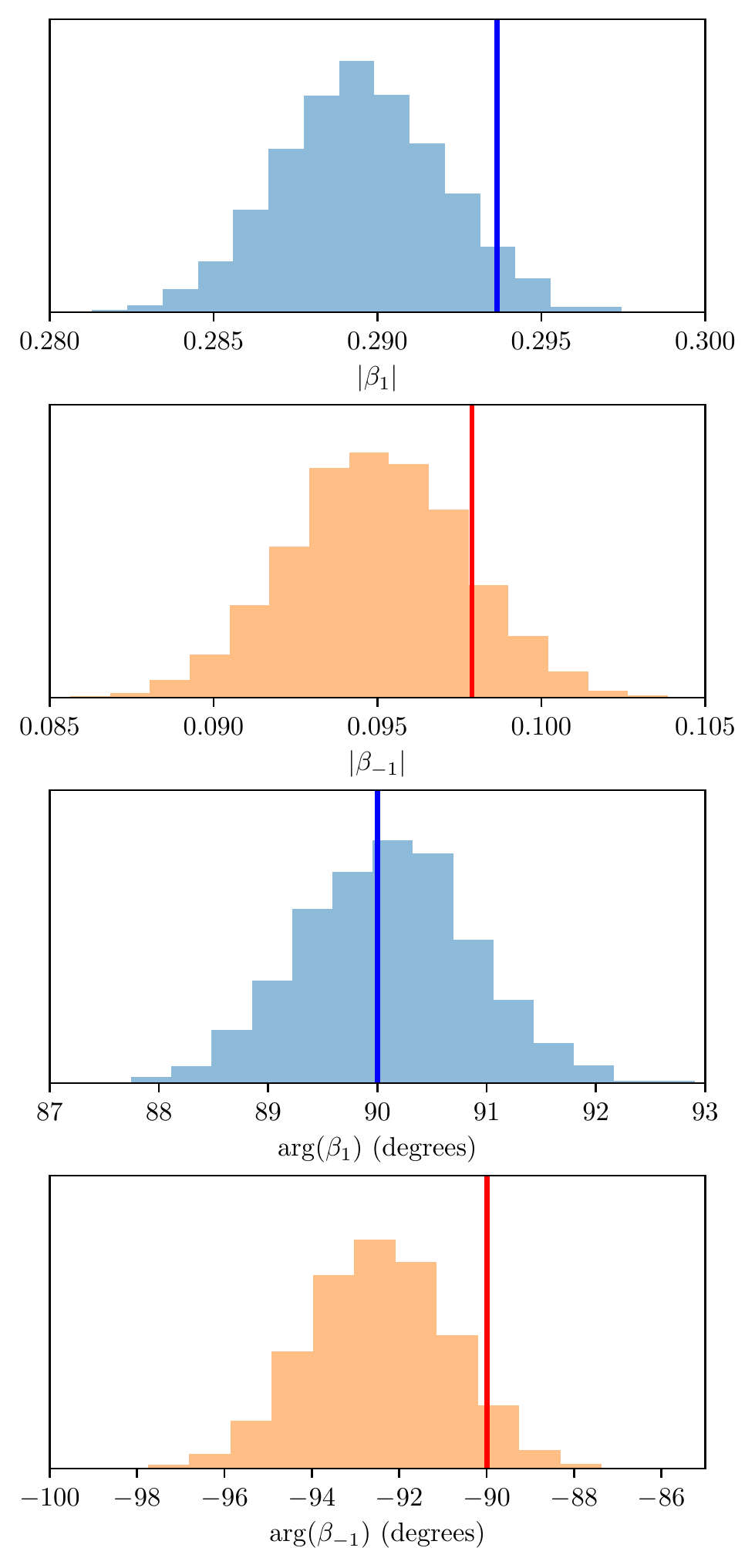}
\caption{Posterior distributions for the $\beta_{1}$ (in blue) and $\beta_{-1}$ (in orange) values corresponding to the \dmc reconstruction of the EHT-like synthetic dataset.  The top two panels show the amplitudes of both quantities, while the bottom two panels show their phases.  In all panels, the value derived from the input image is shown as a vertical line.}
\label{fig:beta_posterior}
\end{figure}

\begin{deluxetable*}{lCCCC}[t]
\tablecolumns{5}
\tablewidth{0pt}
\tablecaption{Station leakages for synthetic data \label{tab:Leakages}}
\tablehead{\colhead{Station} & \colhead{Input $D_R$} & \colhead{Posterior $D_R \pm 1\sigma$} & \colhead{Input $D_L$} & \colhead{Posterior $D_L \pm 1\sigma$}}
\startdata
ALMA     & \hphantom{-}8.0-5.0i & \left( 7.97 - 4.94i \right) \pm \left( 0.14 + 0.11i \right) & \hphantom{-}2.0+7.0i & \left( 2.11 + 7.06i \right) \pm \left( 0.11 + 0.11i \right) \\
APEX     & -6.0+7.0i            & \left( -5.93 + 7.00i \right) \pm \left( 0.11 + 0.12i \right) & -6.0+3.0i & \left( -6.09 + 3.04i \right) \pm \left( 0.12 + 0.10i \right) \\
SMT      & \hphantom{-}4.0-5.0i & \left( 4.02 - 5.03i \right) \pm \left( 0.16 + 0.16i \right) & \hphantom{-}6.0+5.0i & \left( 6.14 + 5.08i \right) \pm \left( 0.19 + 0.18i \right) \\
JCMT     & -4.0+5.0i            & \left( -4.34 + 5.08i \right) \pm \left( 0.35 + 0.37i \right) & -4.0-5.0i & \left( -4.22 - 4.69i \right) \pm \left( 0.33 + 0.33i \right) \\
LMT      & -4.0+3.0i            & \left( -4.07 + 3.08i \right) \pm \left( 0.22 + 0.22i \right) & \hphantom{-}6.0-3.0i & \left( 6.29 - 3.01i \right) \pm \left( 0.23 + 0.26i \right) \\
IRAM 30m & \hphantom{-}2.0+3.0i & \left( 1.75 + 2.13i \right) \pm \left( 0.43 + 0.43i \right) & -2.0-7.0i & \left( -1.63 - 6.97i \right) \pm \left( 0.43 + 0.41i \right) \\
SMA      & -8.0+1.0i            & \left( -8.23 + 1.01i \right) \pm \left( 0.34 + 0.40i \right) & -2.0+1.0i & \left( -2.02 + 1.48i \right) \pm \left( 0.29 + 0.32i \right) \\
\enddata
\tablecomments{A list of the leakage values derived from the \dmc fit to the EHT-like synthetic dataset, compared against the input values for all stations.  The second and fourth columns list the the input values, while the third and fifth columns list the posterior values for the right- and left-hand leakage terms, respectively.  The leakage values in the third and fifth columns are quoted as posterior means, with the posterior standard deviation in both the real and imaginary components following the $\pm$ symbol.  All leakage values are quoted as percentages.}
\end{deluxetable*}

\begin{figure*}
\centering
\includegraphics[width=\textwidth]{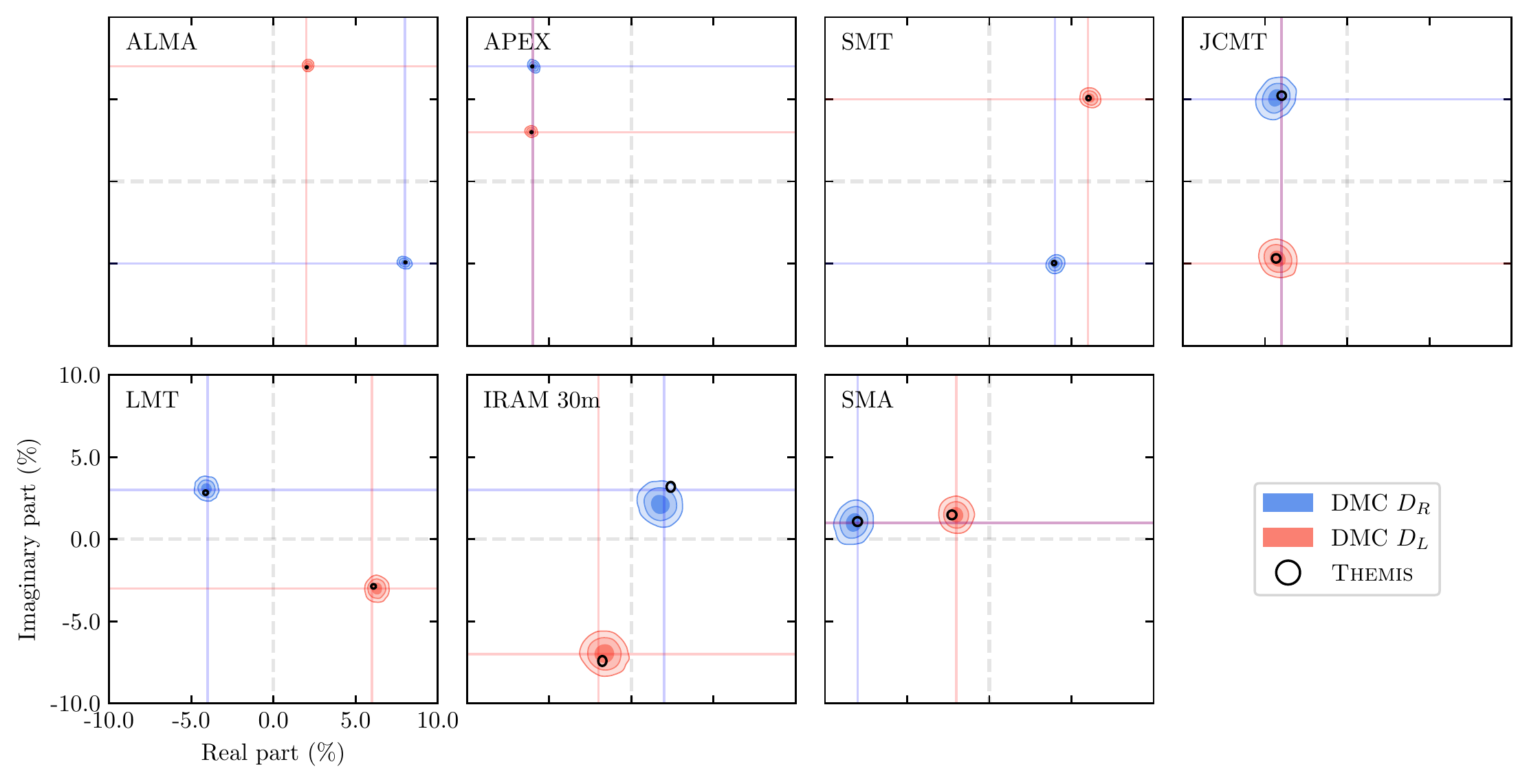}
\caption{Leakage posteriors for the \dmc fit to the EHT-like synthetic dataset described in \autoref{sec:SyntheticData}; each panel shows the result for an individual station, and ground-truth input values are marked using solid vertical and horizontal lines.  The same axis ranges are used for all panels and are explicitly labeled in the bottom-left panel.  Blue contours show the right-hand leakage posteriors, and red contours show the left-hand leakage posteriors.  In all panels, the plotted contours enclose 50\%, 90\%, and 99\% of the posterior probability.  For comparison, we have overplotted the leakage posteriors from \themis as black ellipses, with the sizes of the ellipses along each axis corresponding to the posterior standard deviation.}
\label{fig:dterms_EHT}
\end{figure*}

\subsection{Imaging synthetic data} \label{sec:SyntheticDataImaging}

We use \dmc to image the synthetic dataset described in the previous section, setting $\text{FOV}_x = \text{FOV}_y =60$\,\uas, $N_x = N_y = 30$, $\breve{x}_0 = \breve{y}_0 = 0$\,\uas, and $\Sigma = 6$\,\uas.  The initial total flux estimate $\breve{F}$ is set equal to the visibility amplitude on the shortest baseline, and the gain amplitude prior standard deviations are fixed to $\breve{\sigma}_{R} = \breve{\sigma}_{L} = 0.1$ for all stations.  We select the ALMA station to be our reference antenna, both because it is present at every timestamp and because it provides an absolute EVPA standard \citep{Paper3,Goddi_2019}.\footnote{We note that for this synthetic dataset, all of the stations are effectively calibrated in absolute EVPA.  For real EHT data only the ALMA station has this property, which it inherits from application of the \texttt{PolConvert} procedure \citep{MartiVidal_2016} that gets applied after correlation to convert the mixed-basis polarization products on ALMA baselines to the circular polarization products used by the rest of the array \citep{Paper3,Goddi_2019}.}

\dmc is able to find good fits to this dataset, with reduced-$\chi^2$ values near unity and an additional systematic noise parameter consistent with zero (i.e., the model is able to describe the data to within thermal errors without requiring an additional source of systematic uncertainty).  The image-integrated linear polarization fraction is recovered as $9.96 \pm 0.03$\% (compared to an input value of 10\%), while the image-integrated circular polarization fraction is found to be $1.3 \pm 1.0$\% (compared to an input value of 2\%).

In \autoref{fig:synthetic_DMC}, we show maps of both the mean and the standard deviation of the image ensemble output from \dmc\,-- where each image in the ensemble corresponds to a single sample from the posterior distribution -- and we compare these against the input images of all four Stokes parameters.  Because \dmc produces an estimate of the posterior distribution for every image pixel, arbitrarily complicated single- or multi-point statistics may be computed that self-consistently capture the uncertainties and correlations between pixels.  For instance, image-domain feature extraction techniques -- such as those employed in \cite{Paper4,Paper6} to measure the properties (e.g., diameter, width, orientation) of the ring-like structure seen in M87 -- can be applied to the ensemble of images in the \dmc posterior to yield corresponding distributions for any derived values.

We demonstrate an example of this latter capability in \autoref{fig:beta_posterior}, which shows the derived posteriors for the \citetalias{Palumbo_2020} $\beta_{1}$ and $\beta_{-1}$ parameters.  In this paper, we adopt

\begin{equation}
\beta_m = \frac{1}{F} \iint \left( Q + i U \right) e^{- i m \varphi} r dr d\varphi
\end{equation}

\noindent as our working definition for $\beta_m$, with $F$ the total image flux density and the integral taken over the entire range of image-domain spatial coordinates $(r,\varphi)$.  The \citetalias{Palumbo_2020} $\beta_m$ decomposition permits a quantitative characterization of the polarization structure on a ring, and for the input synthetic dataset only $\beta_{1}$ and $\beta_{-1}$ modes are nonzero.  As shown in \autoref{fig:beta_posterior}, we find that the \dmc fits return posteriors on these two modes that are consistent with the input values.

Furthermore, because \dmc simultaneously models both the image structure and the station-based corruptions, we are also able to investigate the posterior behavior of the latter.  Of most interest for polarimetric reconstructions are the leakage terms, which we list in \autoref{tab:Leakages}.  \autoref{fig:dterms_EHT} shows the leakage posterior distributions for each station in the array.  We see that \dmc is able to recover the correct leakages for all stations and to determine which stations are most well-constrained.

We also compare the \dmc results with those of \themis \citep{Broderick_2020a}, an independent code capable of producing simultaneous image and leakage posteriors for VLBI datasets \citep{Broderick_2021}.  In \autoref{fig:dterms_EHT} and \autoref{fig:EHT_pol_comparison}, the \dmc posteriors are compared against those from \themis for the station leakages and image structure, respectively.  Although the two softwares employ distinct model specifications and exploration schemes, we find that the posterior reconstructions from both codes nevertheless show good agreement.  The \themis fit typically shows only small ($\lesssim$1$\sigma$) shifts in posterior mean relative to the \dmc fit, and the systematically wider posterior widths from the \dmc fit stem from the more permissive station gain priors (\dmc permits the right- and left-hand station gains to vary independently, while \themis enforces equality between both hands).

\begin{figure}
\centering
\includegraphics[width=\columnwidth]{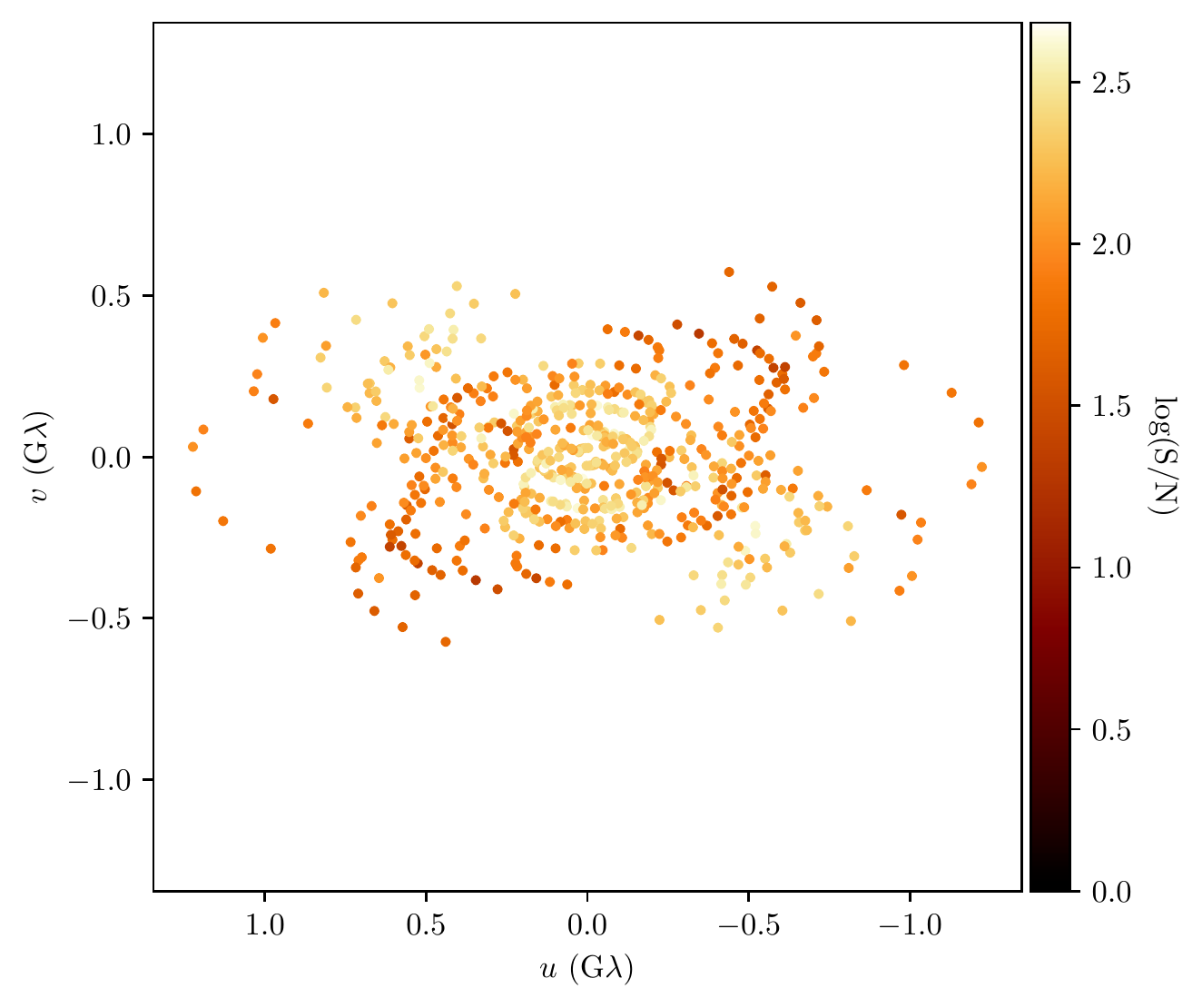}
\caption{Same as \autoref{fig:uv_coverage}, but showing the $(u,v)$-coverage for the OJ 287 dataset used in \autoref{sec:RealDataFits}.}
\label{fig:uv_coverage_OJ287}
\end{figure}

\begin{figure}
\centering
\includegraphics[width=\columnwidth]{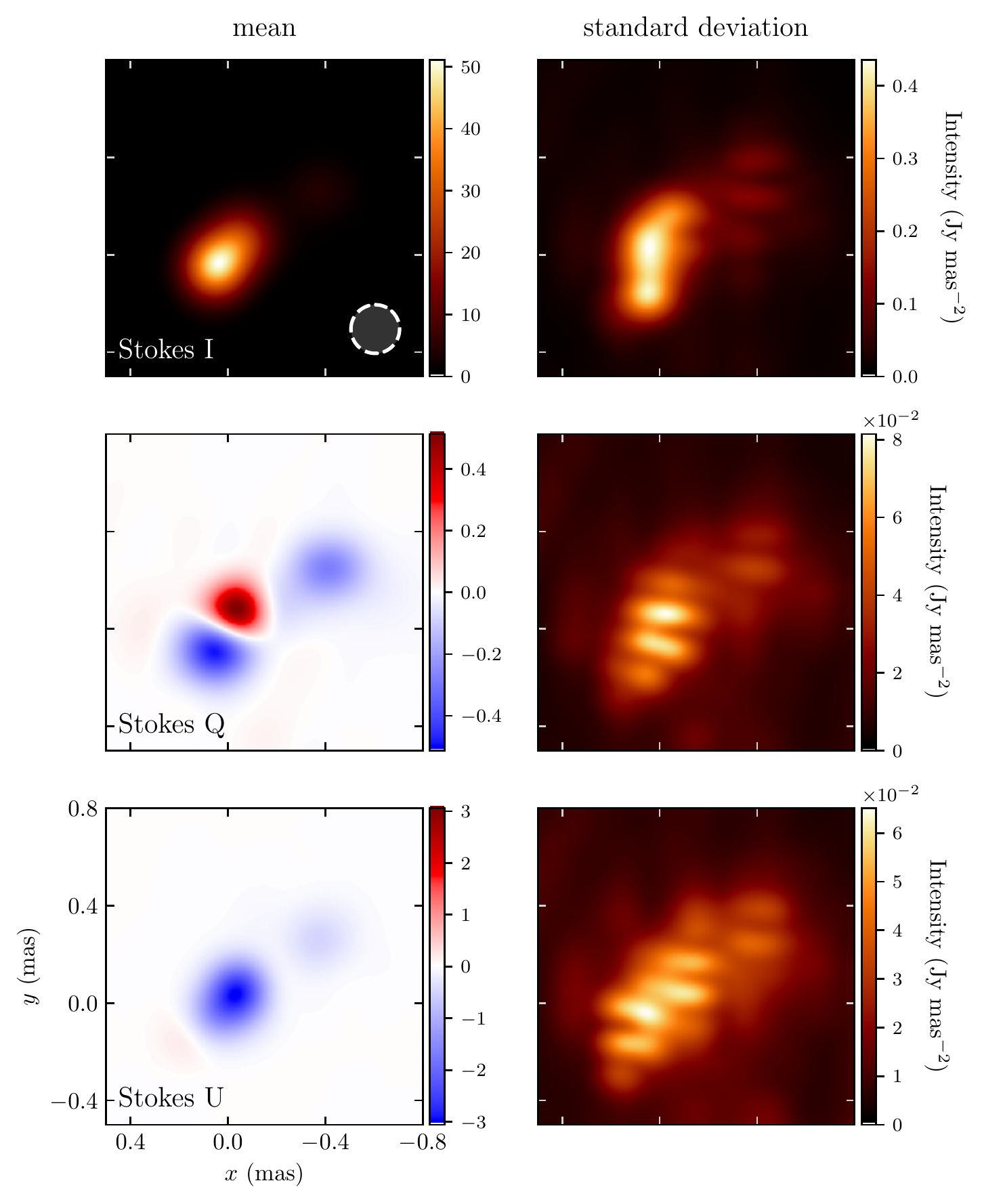}
\caption{Similar to \autoref{fig:synthetic_DMC}, but for the \dmc fit to the OJ 287 dataset (for which we have no ground-truth image).  No significant Stokes V signal is detected in this dataset, so we show here only the Stokes I, Q, and U maps.  The 0.2\,mas FWHM of the Gaussian smoothing kernel is shown in the lower right-hand corner of the top left panel, and we note again that this kernel does not represent a ``restoring beam'' but rather a convolving function that is self-consistently applied during imaging.}
\label{fig:OJ287_DMC}
\end{figure}

\subsection{Imaging real data} \label{sec:RealDataFits}

We also apply \dmc to a real dataset, for which we use a polarimetric VLBI observation of the blazar OJ 287 obtained from the Boston University Blazar Research Group \citep{Jorstad_2016,Jorstad_2017}.  A fully calibrated version of this dataset is publicly available\footnote{\url{http://www.bu.edu/blazars/VLBAproject.html}}, but we wish to demonstrate \dmc's ability to perform calibration itself alongside image reconstruction.  Furthermore, applying \dmc to a pre-calibrated dataset may violate our assumption that the leakage terms are constant in time (see \autoref{app:ResidualJones}).  In this paper we thus use an earlier version of the OJ 287 dataset for which only the a priori amplitude and phase calibrations have been applied (S. Jorstad 2020, private communication), but for which the full set of gain and leakage corruptions have not yet been removed. The observation was carried out with the Very Long Baseline Array (VLBA) on January 3, 2020 at an observing frequency of 43\,GHz, and the data reduction procedure is described in \cite{Jorstad_2005}.  The original dataset is split into four separate frequency bands, but we combine bands during imaging to produce a single image and set of leakage terms.  To reduce data volume, we coherently average the visibilities on a per-scan basis prior to imaging; the $(u,v)$-coverage of the resulting dataset is shown in \autoref{fig:uv_coverage_OJ287}.

We run \dmc on the OJ 287 dataset using $\text{FOV}_x = \text{FOV}_y =2$\,mas, $N_x = N_y = 20$, $\breve{x}_0 = \breve{y}_0 = 0$\,mas, and $\Sigma = 0.2$\,mas.  As with the synthetic dataset, we take the initial total flux estimate $\breve{F}$ to be equal to the visibility amplitude on the shortest baseline.  The gain amplitude prior standard deviations are fixed to $\breve{\sigma}_{R} = \breve{\sigma}_{L} = 0.05$ for all stations, and we select the Pie Town (PT) station as our reference antenna.

\begin{figure*}
\centering
\includegraphics[width=\textwidth]{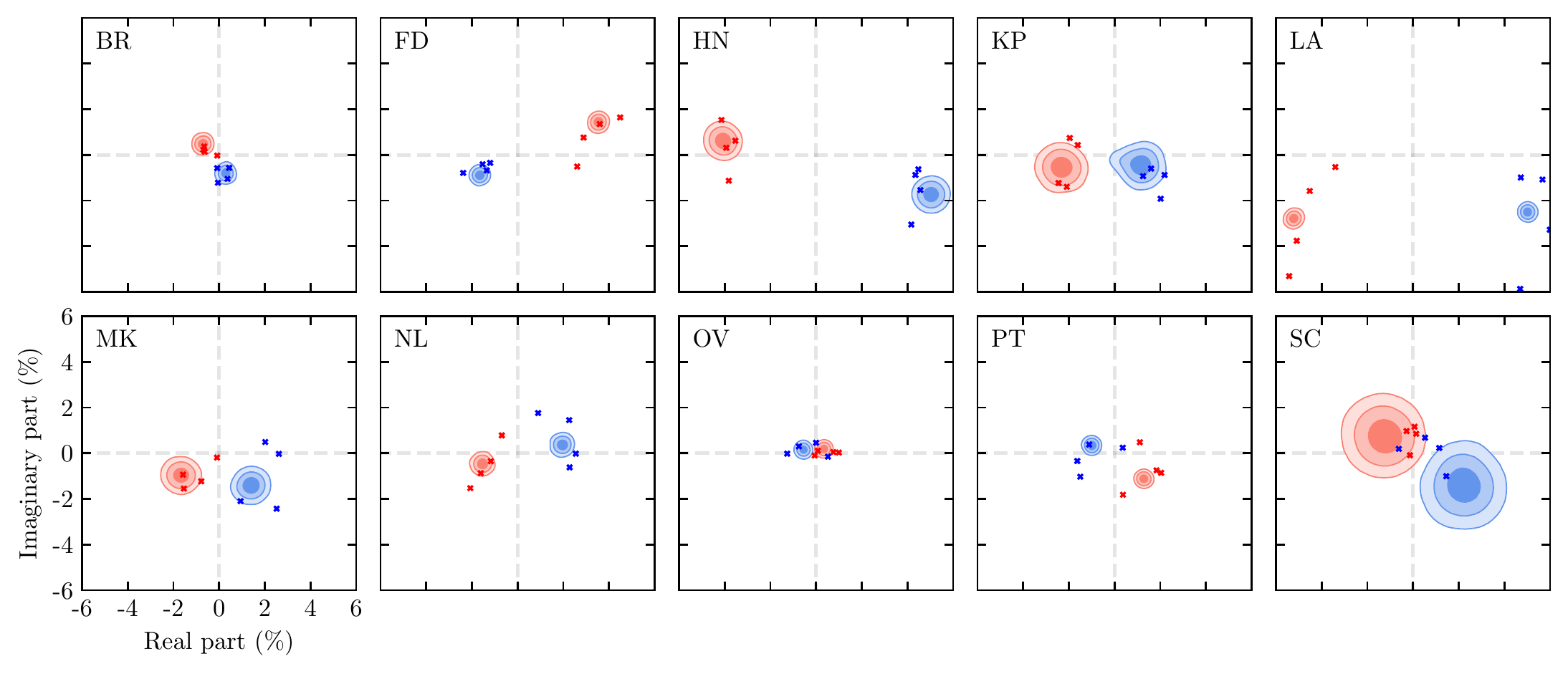}
\caption{Similar to \autoref{fig:dterms_EHT}, but for the \dmc fit to the OJ 287 dataset; as in \autoref{fig:dterms_EHT}, the blue contours correspond to right-hand leakages and the red contours to left-hand leakages.  The overplotted blue and red crosses indicate the right and left leakage solutions, respectively, derived from the same dataset using CLEAN and LPCAL (S. Jorstad 2020, private communication).  There are four crosses associated with each leakage term because the original CLEAN analysis imaged each of the four frequency bands separately, while for the current \dmc analysis we image all bands simultaneously and solve for a single set of leakages.}
\label{fig:dterms_OJ287}
\end{figure*}

\begin{figure*}
\centering
\includegraphics[width=0.8\textwidth]{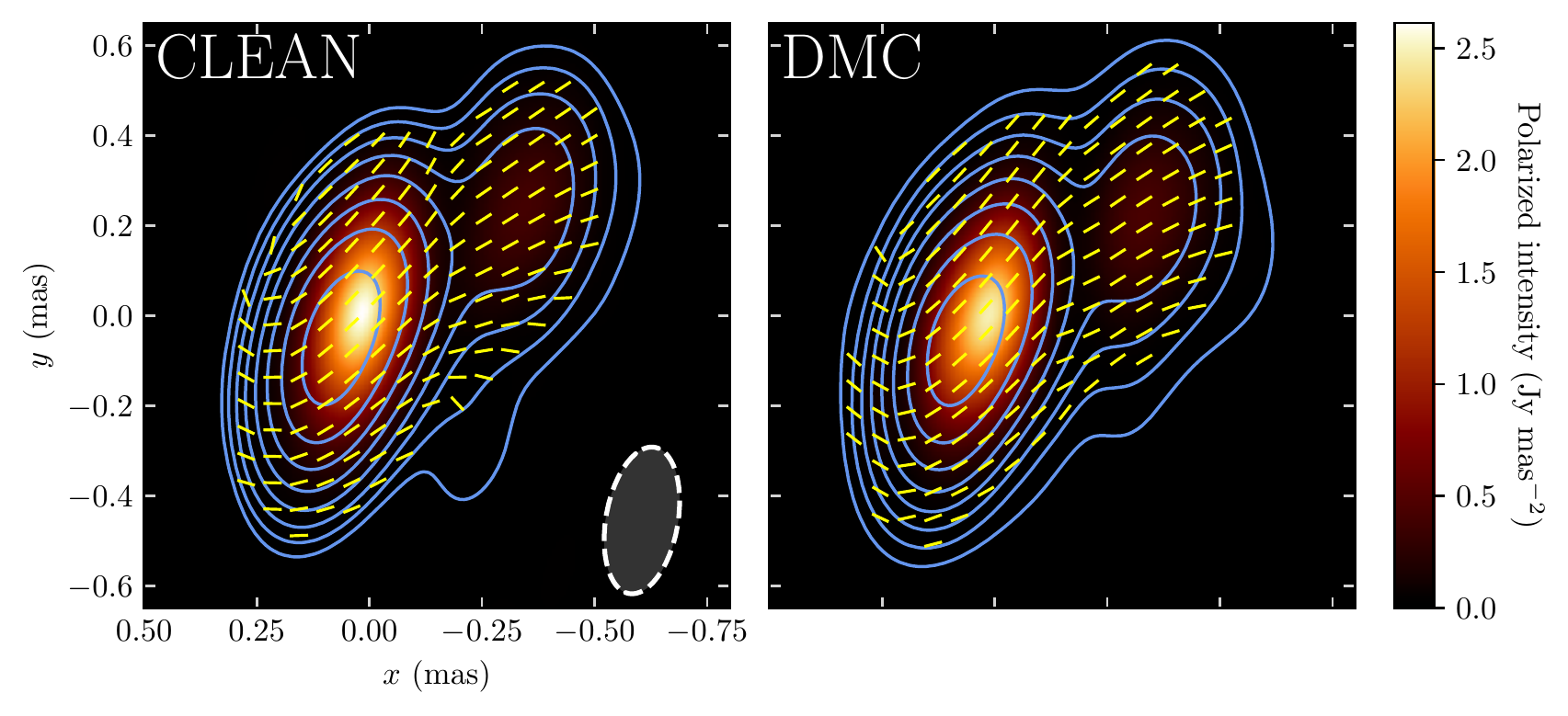}
\caption{Analogous to \autoref{fig:EHT_pol_comparison}, but showing a comparison of the linear polarization structure in the OJ 287 dataset as recovered using CLEAN (left) and \dmc (right).  The outermost Stokes I contour levels start at 0.5\% of the peak intensity, and the contoured value increases inwards by factors of 2.  Both images have been convolved with the CLEAN beam (dimensions $0.33 \times 0.16$\,mas, position angle $-10^{\circ}$ East of North) shown in the lower right-hand corner of the left panel.}
\label{fig:OJ287_pol_comparison}
\end{figure*}

\dmc is able to find good fits to this dataset, achieving a reduced-$\chi^2$ value near unity without requiring any additional systematic noise; i.e., $f$ from \autoref{eqn:SystematicNoise} has a 90\% confidence upper limit of $f < 0.2$\% and is consistent with being zero.  \autoref{fig:OJ287_DMC} shows our reconstructed images for each Stokes parameter, and \autoref{fig:dterms_OJ287} shows the derived station leakages and compares them against the results from LPCAL.  The LPCAL leakages were determined separately for each of the four frequency bands in the original dataset, and are the result of averaging leakage solutions across 15 different observational targets (including OJ 287).  We find that the leakage posteriors recovered by \dmc -- which are determined by solving for only a single leakage term for all four frequency band at once -- behave similarly to the LPCAL values for all stations and are largely consistent with the average of the LPCAL values across bands.

\autoref{fig:OJ287_pol_comparison} shows a comparison between the \dmc and CLEAN image reconstructions for the OJ 287 dataset, after restoring both with the CLEAN beam.  We see good agreement between both the Stokes I and linearly polarized image structures across most of the image, with noticeable deviations manifesting only at the $\lesssim$1\% Stokes I contour level (which, as can be seen in \autoref{fig:OJ287_DMC}, is at the level of the uncertainty in the \dmc Stokes I map).

\section{Summary} \label{sec:Summary}

In this paper we have presented \dmc, a Python-based software package for performing polarimetric imaging of VLBI data.  \dmc simultaneously reconstructs the full-Stokes image structure and solves for the station-based gain and leakage calibration terms within a probabilistic formalism.  The output of \dmc is a sample of parameter values drawn from the posterior distribution, such that instead of a single image and associated set of calibration quantities one obtains an ensemble of images and sets of calibration quantities.  \dmc explores this posterior space using the NUTS HMC sampler implemented within PyMC3.  Using this posterior distribution it is possible to determine not only a best-fit value but also a self-consistent measure of uncertainty in any of the modeled parameters, including the sky-plane emission structure in all four Stokes parameters as well as the aforementioned calibration quantities.  \dmc therefore provides a tool for combining the information across different regions of an image (e.g., for averaging low-flux regions) or across multiple images (e.g., for spectral index or Faraday rotation analyses) in a manner that properly accounts for the associated uncertainties and correlations.

We have demonstrated the effectiveness of \dmc on both synthetic and real data.  The results from running \dmc on an EHT-like synthetic dataset show that \dmc is capable of accurately recovering both the image structure and the ground-truth calibration quantities.  Our results are compatible with those of \themis, which is also capable of exploring the full posterior space and returns distributions that are consistent with \dmc's.  Similarly, we find that running \dmc on a real VLBA dataset recovers an image structure and leakage terms that agree well with those derived from an independent analysis using CLEAN and LPCAL.

\dmc is built on a highly flexible modeling framework, and it continues to be developed.  Future \dmc capabilities may include more sophisticated priors on both the image and calibration terms, a more physically-motivated calibration model, hybrid imaging (i.e., simultaneous imaging and modeling), compound sampling options, and time-, frequency-, and scale-dependent imaging.

\acknowledgments
I thank the anonymous referee for a constructive report that improved the quality of this paper.  I am grateful to Lindy Blackburn, Avery Broderick, Andrew Chael, Shep Doeleman, Michael Johnson, Iv\'an Mart\'i-Vidal, and Daniel Palumbo for helpful discussions and comments.  I'd like to further thank Andrew Chael for his guidance on generating synthetic data using the \texttt{eht-imaging} library, Avery Broderick for providing \themis results, and Svetlana Jorstad and Alan Marscher for providing the uncalibrated dataset, LPCAL leakage solutions, and polarized CLEAN images of OJ 287.  Support for this work was provided by the NSF through grants AST-1952099, AST-1935980, AST-1828513, and AST-1440254, and by the Gordon and Betty Moore Foundation through grant GBMF-5278.  This work has been supported in part by the Black Hole Initiative at Harvard University, which is funded by grants from the John Templeton Foundation and the Gordon and Betty Moore Foundation to Harvard University.
This study makes use of 43 GHz VLBA data from the VLBA-BU Blazar Monitoring Program (VLBA-BU-BLAZAR;
\url{http://www.bu.edu/blazars/VLBAproject.html}), funded by NASA through the Fermi Guest Investigator Program. The VLBA is an instrument of the National Radio Astronomy Observatory. The National Radio Astronomy Observatory is a facility of the National Science Foundation operated by Associated Universities, Inc. \\

\software{\texttt{eht-imaging} \citep{Chael_2016,Chael_2018}, \texttt{ehtplot}\footnote{\url{https://github.com/liamedeiros/ehtplot}}, PyMC3 \citep{Salvatier_2016}, \themis\citep{Broderick_2020a,Broderick_2021}}

\clearpage

\bibliography{references}{}
\bibliographystyle{aasjournal}

\appendix

\section{Residual Jones matrices} \label{app:ResidualJones}

In this section we explore the notion of a ``residual'' Jones matrix; i.e., what is the form of the Jones matrix for the remaining calibrations after an imperfect calibration has been applied?  Consider a dataset that has had such an attempted (but imperfect) calibration applied to it.  The true Jones matrix $\textbf{J}$ looks like (see \autoref{eqn:JonesMatrix})

\begin{equation}
\textbf{J} = \begin{pmatrix}
G_R & 0 \\
0 & G_L
\end{pmatrix} \begin{pmatrix}
1 & D_R \\
D_L & 1
\end{pmatrix} \begin{pmatrix}
e^{-i \phi} & 0 \\
0 & e^{i \phi}
\end{pmatrix} ,
\end{equation}

\noindent but the applied inverse Jones matrix $\hat{\textbf{J}}$ has some imperfections in it.  We can assume that these imperfections manifest in the station gain and leakage terms, but not in the field rotation angles.  $\hat{\textbf{J}}$ may thus be written as

\begin{equation}
\hat{\textbf{J}} = \begin{pmatrix}
G_R \left( 1 + \epsilon_{R} \right) & 0 \\
0 & G_L \left( 1 + \epsilon_{L} \right)
\end{pmatrix} \begin{pmatrix}
1 & D_R \left( 1 + \delta_{R} \right) \\
D_L \left( 1 + \delta_{L} \right) & 1
\end{pmatrix} \begin{pmatrix}
e^{-i \phi} & 0 \\
0 & e^{i \phi}
\end{pmatrix} .
\end{equation}

\noindent The act of applying the imperfect calibration corresponds to attempting to ``invert'' $\textbf{J}$ using $\hat{\textbf{J}}^{-1}$, which yields a residual Jones matrix,

\begin{eqnarray}
\textbf{R} & \equiv & \hat{\textbf{J}}^{-1} \textbf{J} \nonumber \\
& = & \begin{pmatrix}
\Gamma_R & 0 \\
0 & \Gamma_L
\end{pmatrix} \begin{pmatrix}
1 & \Delta_R \\
\Delta_L & 1
\end{pmatrix} \begin{pmatrix}
e^{-i \phi} & 0 \\
0 & e^{i \phi}
\end{pmatrix} .
\end{eqnarray}

\noindent Here, we have defined the residual gains,

\begin{subequations}
\begin{equation}
\Gamma_R \equiv e^{i \phi} \left( \frac{r D_R D_L \left( 1 + \delta_{R} \right) - 1}{\left[ D_R \left( 1 + \delta_{R} \right) D_L \left( 1 + \delta_{L} \right) - 1 \right] \left( 1 + \epsilon_{R} \right)} \right) ,
\end{equation}
\begin{equation}
\Gamma_L \equiv e^{-i \phi} \left( \frac{D_R D_L \left( 1 + \delta_{L} \right) - r}{\left[ D_R \left( 1 + \delta_{R} \right) D_L \left( 1 + \delta_{L} \right) - 1 \right] \left( 1 + \epsilon_{R} \right)} \right) ;
\end{equation}
\end{subequations}

\noindent the residual leakages,

\begin{subequations}
\begin{equation}
\Delta_R \equiv D_R \left( \frac{1 + \delta_{R} - r}{D_R D_L \left( 1 + \delta_{R} \right) - r} \right) ,
\end{equation}
\begin{equation}
\Delta_L \equiv D_L \left( \frac{r \left( 1 + \delta_{L} \right) - 1}{r D_R D_L \left( 1 + \delta_{L} \right) - 1} \right) ;
\end{equation}
\end{subequations}

\noindent and the fractional gain residuals,

\begin{equation}
r \equiv \frac{1 + \epsilon_{L}}{1 + \epsilon_{R}} .
\end{equation}

\noindent We see that while the residual Jones matrix $\textbf{R}$ can be cast in the form of the original Jones matrix, the residual gain and leakage terms may no longer adhere to the same set of assumptions that apply to the true gains and leakages.  In particular, the residual leakages depend on the fractional gain residuals, $r$, which may not be constant in time.  It may therefore be preferable to redo calibration entirely rather than attempting to incrementally calibrate a partially-calibrated dataset; in \autoref{sec:RealDataFits}, we pursue the former strategy.

\end{document}